# Additives That Prevent Or Reverse Cathode Aging In Drift Chambers With Helium-Isobutane Gas*

A. M. Boyarski[1]

*Stanford Linear Accelerator Center, Stanford, CA 94309, USA*

**Abstract**

Noise and Malter breakdown have been studied at high rates in a test chamber having the same cell structure and gas as in the BaBar drift chamber. The chamber was first damaged by exposing it to a high source level at an elevated high voltage, until its operating current at normal voltages was only ~0.5nA/cm. Additives such as water or alcohol allowed the damaged chamber to operate at 25 nA/cm, but when the additive was removed the operating point reverted to the original low value. However with 0.02% to 0.05% oxygen or 5% carbon dioxide the chamber could operate at more than 25 nA/cm, and continued to operate at this level even after the additive was removed. This shows for the first time that running with an $O_2$ or $CO_2$ additive at high ionisation levels can cure a damaged chamber from breakdown problems.

## 1. Introduction

Many experiments have found that drift chambers operating in high ionisation environments become noisy and experience high voltage trips. Adding various alcohols [1][2] or water vapor (see Va'Vra review article [3] for list of detectors) to the chamber gas has been shown to alleviate this problem in a number of chambers. Most of these chambers use an argon-based gas mixture. There is less experience with helium-based gas mixtures, as used in BaBar.

Initially, the BaBar drift chamber used a gas mixture of helium (80%) and isobutane (20%). During the turn on at PEPII, the backgrounds were high and the chamber soon developed current spikes and increased wire currents. When 0.35% water vapor was added to the chamber gas the current spikes ceased and the currents returned to normal. The BaBar drift chamber operates at a sense wire current of 0.3nA/cm at the present time, but future improvements in PEPII will increase the current.

* Work supported by Department of Energy contract DE-AC03-76SF00515.

[1] Presented at the Aging Phenomena In Gaseous Detectors workshop at DESY, Germany, Oct 2-5, 2001. To be published in Nuclear Instruments and Methods, section A.

Since the long term effects of a water additive at high chamber currents is not well known, a test chamber was built to study this as well as other additives at high ionisation levels.

Alcohol additives have reduced chamber currents in argon based gas mixtures. It is believed that $CO_2$ has prolonged the chamber life in some experiments [4], most likely due to its oxygen component that reacts with carbon-based deposits on the wires. If the latter is true then oxygen alone should also be beneficial. Va'Vra [3] points out that plasma chemists find that oxygen helps to remove hydrocarbon radicals from cathodes, and that oxygen is believed to have stabilized operation in two drift chamber cases.

In view of the above, the additives $H_2O$, Methylal, 2-Propanol, $CO_2$, and $O_2$ were studied in a helium-based gas at high ionisation levels and the results are reported here.

## 2. Apparatus

A small test chamber was built having one hexagonal BaBar-like cell as shown in Fig. 1. The sense wire (20um gold coated tungsten-rhenium) is surrounded by six field wires (120um gold coated aluminum) at 1 cm wire spacing. The six outer bias wires at 1300 volts together with 2050 volts on the sense wire provide a 19.2 KV/cm field on the field (cathode) wires and 239 KV/cm on the sense (anode) wire.

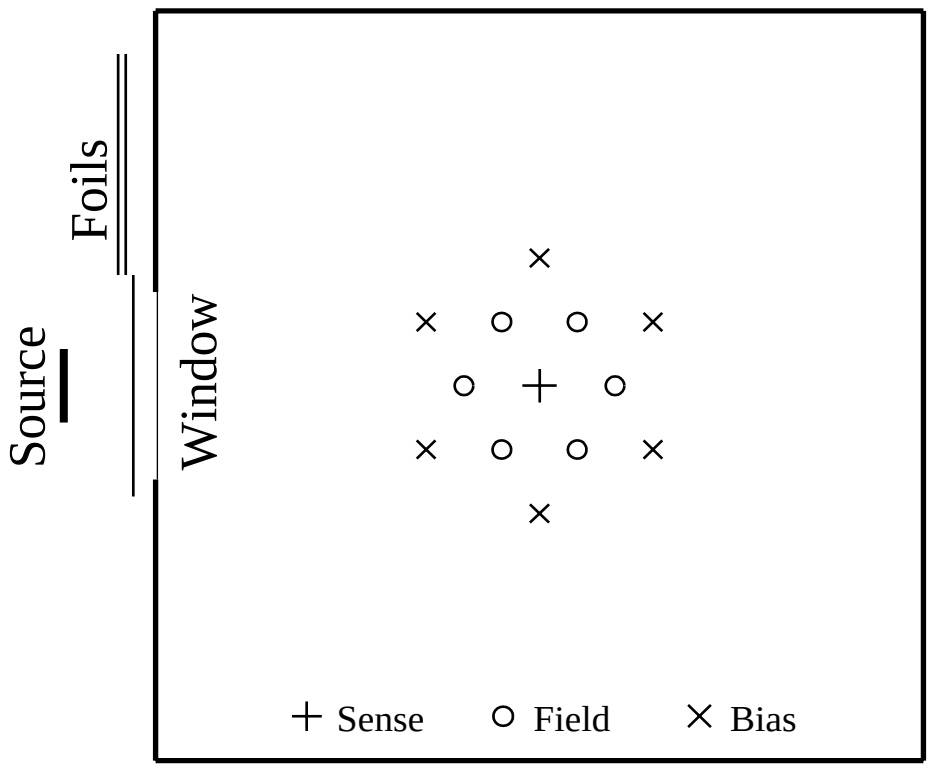

Figure 1. The test chamber. A single drift cell is enclosed in a square aluminum tube 10.2 cm per side, and 30.5 cm long.

A 100 mCi $Fe^{55}$ source and attenuation foils provide any desired wire current up to 30 nA/cm. The source could be closed with a 6 mm thick aluminum shutter. A thin Mylar window restricted the ionisation to the middle 22 cm of the sense wire length. The maximum current at mid-length is calculated to be (total wire current) /13 (nA/cm).

Mass flow controllers for the helium, isobutane and additive gases set the gas mixture. For liquid additives, a portion of the helium flow was bubbled through the additive. The total flow rate was 125 cc/min, which gave a volume change every 24 minutes. The chamber operated at one atmosphere pressure, and the exhaust gas was vented through a bubbler to the atmosphere.

A Pico ammeter (Keithly 487) and a multi-channel analyzer (Perkin Elmer/Ortec 142PC, 570, and TRUMP PCI-8K) measured the cathode current and anode pulse spectrum respectively. The 8K-channel range in the analyzer could accommodate both the $Fe^{55}$ peak at channel 5500 and the signal from one-electron avalanches at channel ~32. The analyzer did not record pulse heights below channel 20. The top plot in Fig. 2 shows the small pulse spectrum from $Fe^{55}$ at a low source level, while the bottom plot shows the spectrum when incandescent light is allowed into the chamber. The latter spectrum is from photoelectrons off the cathode wires, and establishes the shaded channels (30-120) to be used for counting single electron avalanches. The small pulses from $Fe^{55}$ are due to conversion locations grazing the boundary of the cell with only a single (few) electron(s) reaching the anode.

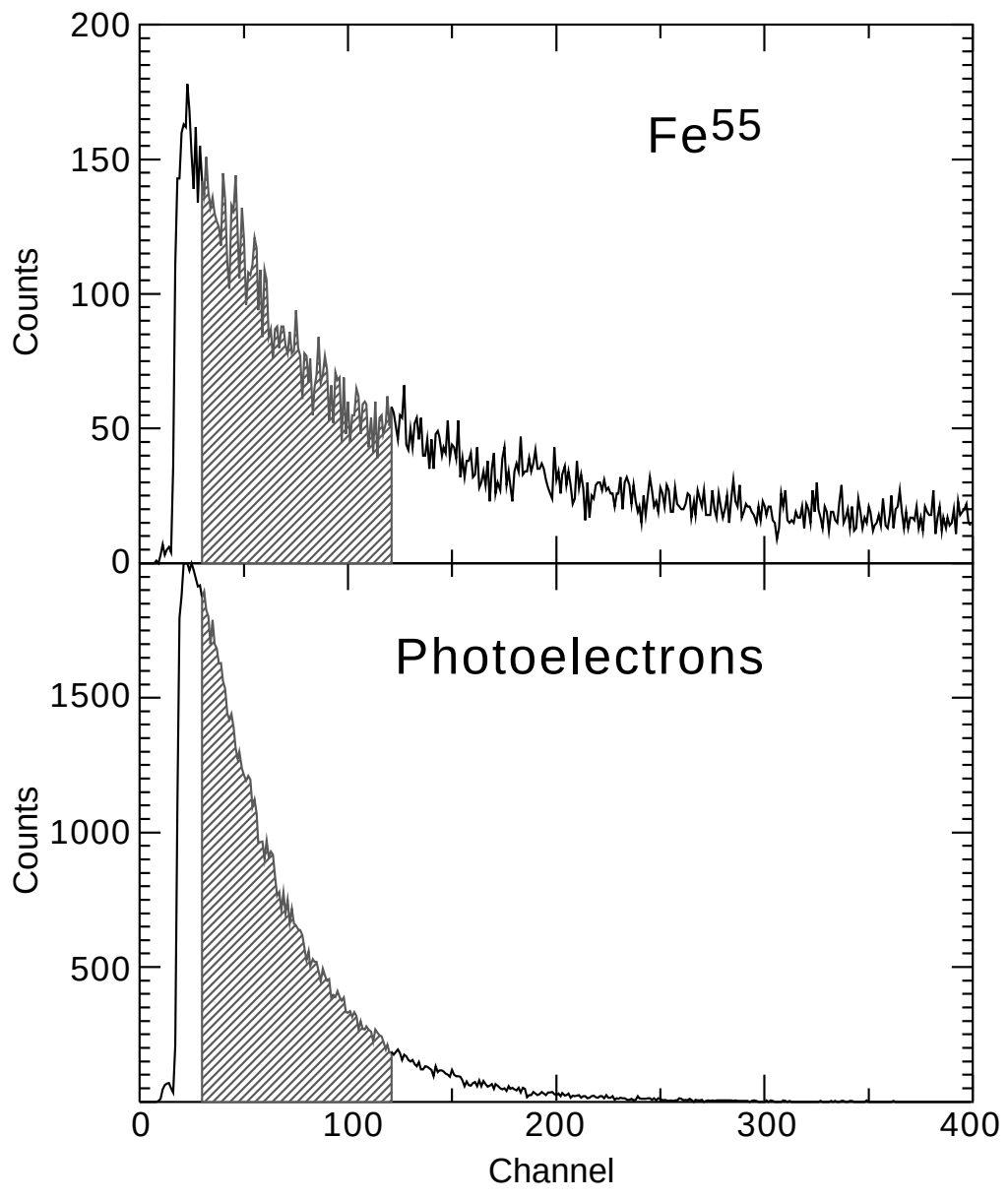

Figure 2. Pulse height spectrum for small pulses from $Fe^{55}$ and from single photoelectrons. The shaded channels are used to count single electron avalanches. The $Fe^{55}$ peak is off scale at channel 5500.

## 3. Measurements

With only the Helium:Isobutane (80:20) gas, the chamber could operate at wire currents ≤1 nA/cm. Beyond this level, the chamber current suddenly jumps by two orders of magnitude, and remains elevated even if the source is removed until the voltage is lowered. This is the Malter effect [5], resulting from electrons that are pulled out of the cathode surface producing a multitude of single-electron avalanches in the chamber.

In order to better understand the onset of the Malter currents, the transient behavior of small pulses was measured in response to a step in chamber ionisation. This was done by closing the source shutter for at least 5 minutes and then opening the shutter while recording pulse spectra every few seconds. The ratio of small pulse counts to peak counts was then plotted versus time. Such plots were recorded for various chamber currents. Fig. 3 shows the transient behavior of smnall pulses for wire currents from 0.25 to 1.3 nA/cm in the Helium:Isobutane (80:20) gas with no additives.

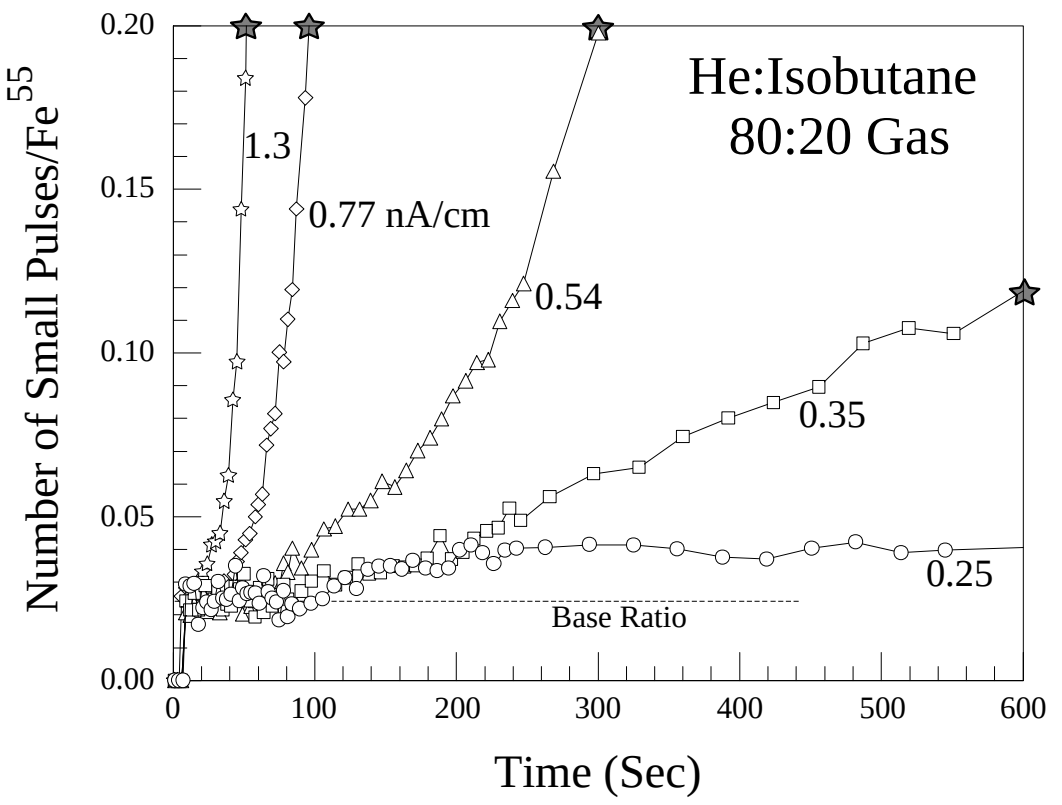

Figure 3. Transient behavior of small pulses in the (80:20) gas after opening the source shutter, for various levels of wire current. A star at a boundary indicates that breakdown occurred outside of the plot area.

When the source is first opened, the small pulse ratio starts at the same 0.025 base value for all current levels. At the highest current (1.3 nA/cm) the ratio grows rapidly until the chamber breaks down after 60 seconds. At lower currents the break down takes longer to develop. At 0.25 nA/cm the chamber does not breakdown although there is an elevation in the number of small pulses.

Some additives allowed chamber operation beyond wire currents of 10 nA/cm. The transient behavior with 2-Propanol, $CH_3CHOHCH_3$, is shown in Fig. 4. At concentrations of 1% and 0.5% there is no small pulse activity for wire currents of 10 nA/cm or more, while at 0.25% there is an initial increase at 12.5 nA/cm and then a gradual decline. The declining

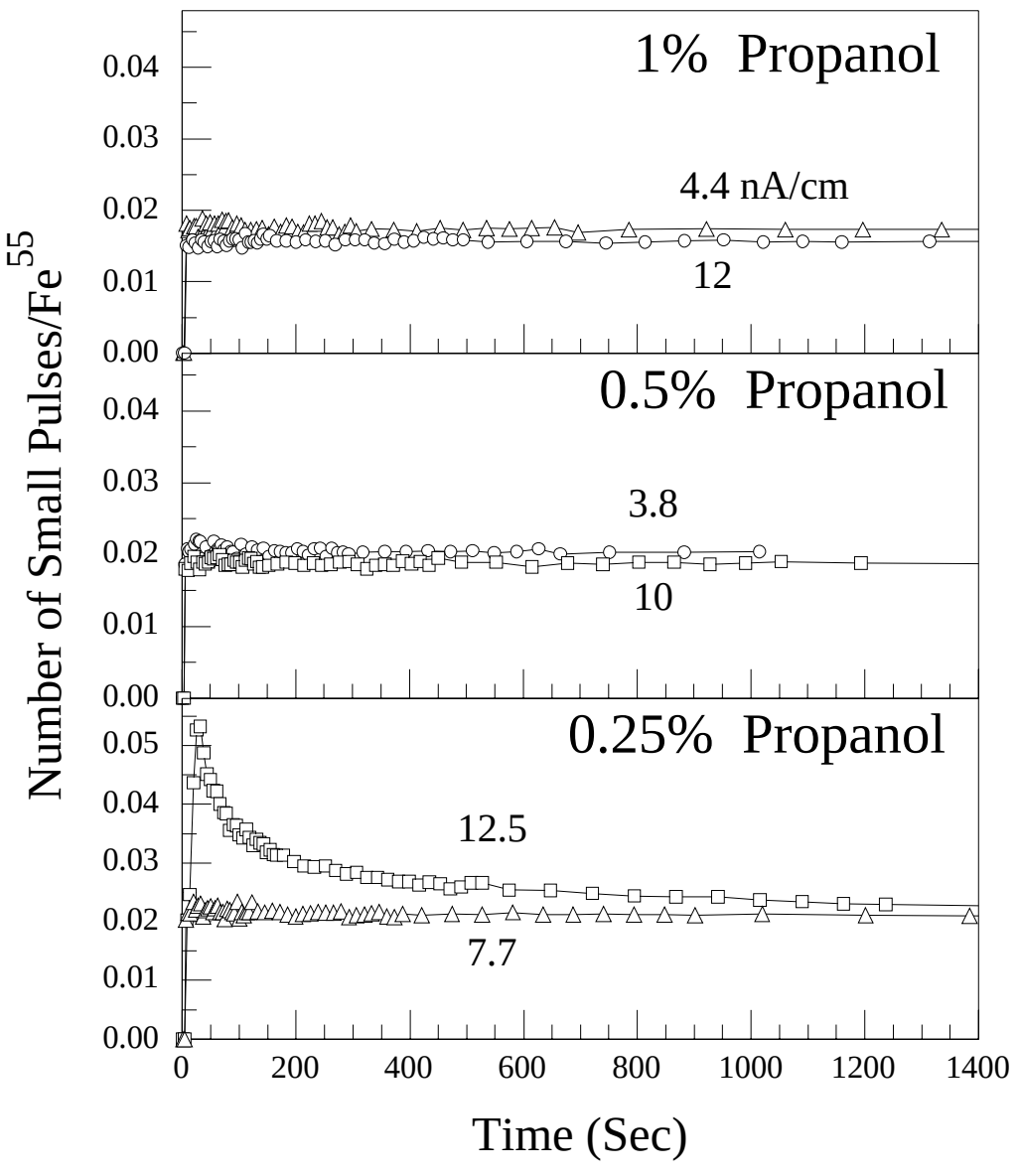

Figure 4. Transient behavior of small pulses with an alcohol additive 2-Propanol. Concentrations above 0.5% show no small pulse activity.

feature is discussed later.

The transient behavior for a water additive is shown in Fig. 5. There is no visible small pulse activity with 0.35% or 0.18% water.

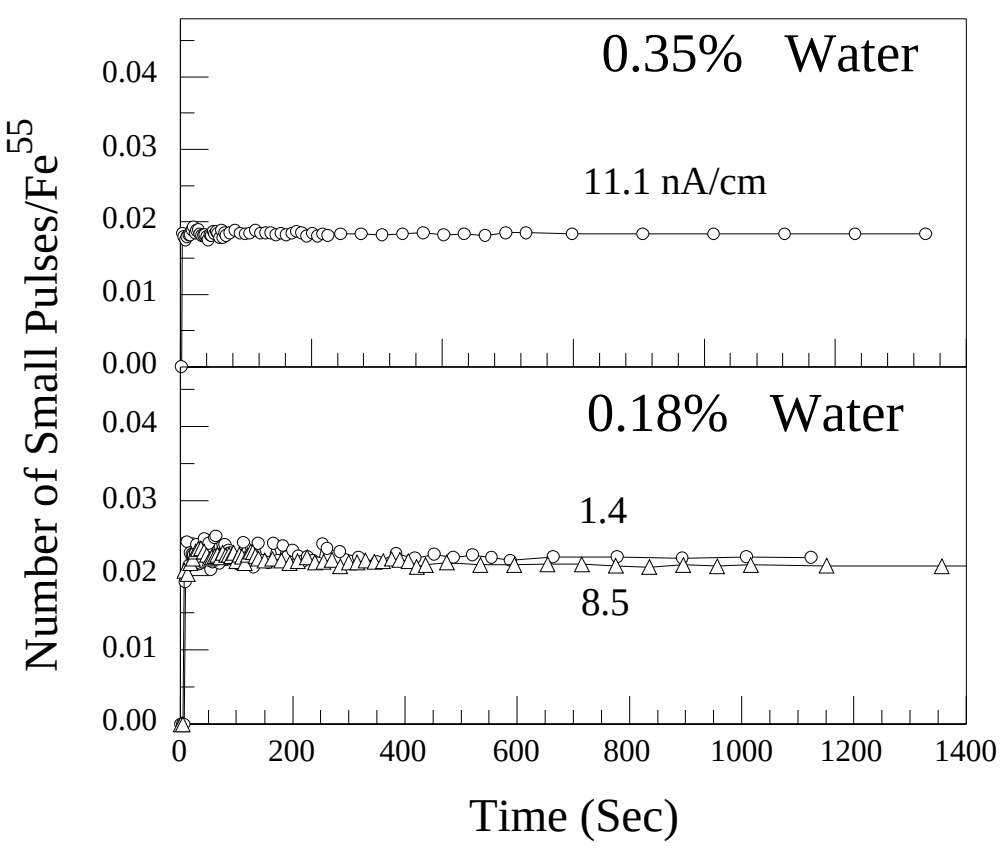

Figure 5. The small pulse transient behavior with a 0.35% or 0.18% water additive shows no increase in small pulses.

A Methylal additive, $CH_2(OCH_3)_2$, was also tried. It showed just a slight amount of activity at a 4% concentration and broke down at a 2% concentration at a lesser wire current.

The final two additives tried in this study, $O_2$ and $CO_2$, behaved differently. Whereas the previous additives allowed immediate operation at higher chamber currents, these did not. With $O_2$ or $CO_2$ it was necessary to increase the ionisation slowly. If brought on too quickly, the chamber would ignite in Malter mode, but after repeated attempts would become stable. Fig. 6 shows the anode current with 0.05% oxygen. After 2 hours of training the maximum $Fe^{55}$ source was reached, at an anode current of 375 nA (29 nA/cm wire current).

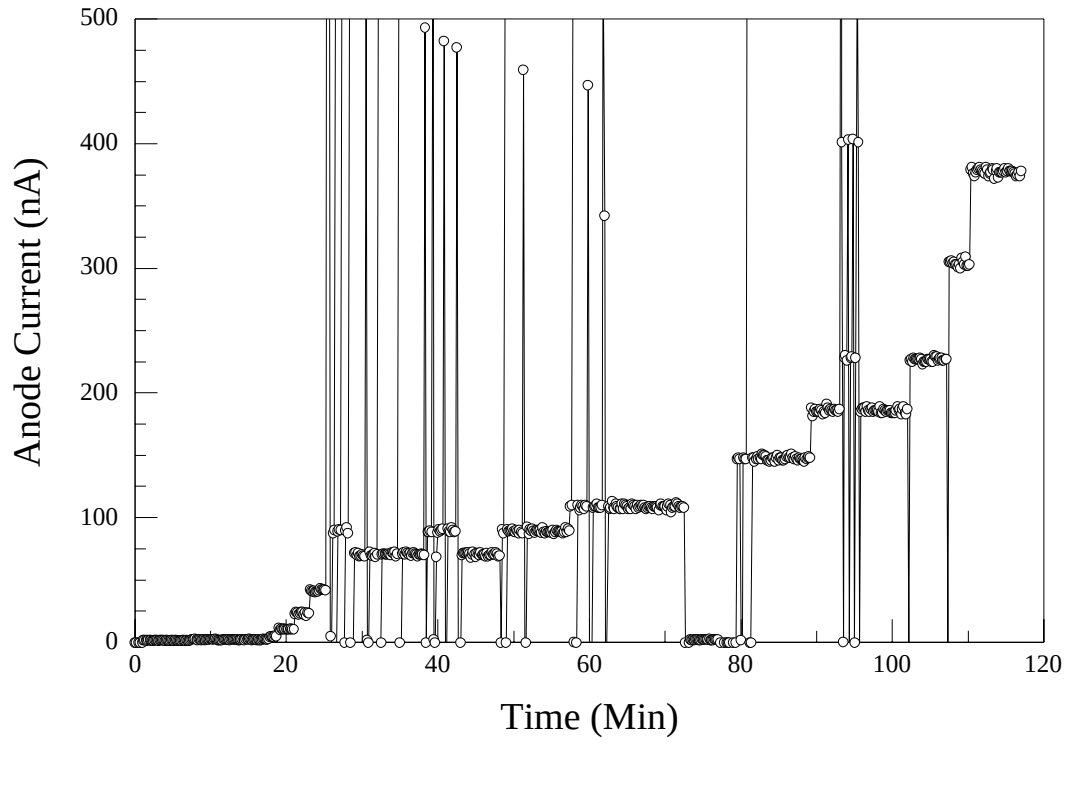

Figure 6. The anode current during a curing cycle with 0.05% $O_2$ is shown. The spikes are chamber trips, but after repeated attempts the chamber reaches a stable state at each current level up to the maximum $Fe^{55}$ source level.

But more importantly, when the $O_2$ was removed from the chamber gas, the chamber could still operate at the maximum 29 nA/cm wire current. The $O_2$ additive was able to cure a damaged chamber. This measurement was repeated several times (by damaging the chamber at high source levels and elevated high voltage, followed by a cure cycle) with the same result.

The fact that the chamber could not operate immediately at high currents with oxygen but had to be slowly trained is indicative that the oxygen reacts with and removes ingredients on the wires over a period of time. It requires both oxygen and a high chamber current to cure the chamber. Most likely, the ionic current passing through the high electric field near the wire surface heats up the local gas, which allows reaction of oxygen with the carbon based deposits (polymers) on the wire surface. The CO, $CO_2$, etc. combustion products are removed from the chamber by the flowing gas.

$CO_2$ showed similar results but the curing time was longer, 35 hours with 5% $CO_2$.

A combination of $H_2O$ and $O_2$ was tried to see if oxygen could cure while the chamber was running with a water additive. It was found that although some curing was seen, it was at a much lower level. After curing for 40 hours, the chamber operated at only 3 nA/cm maximum with the additives removed.

Table 1 summarizes all the measurements.

Table 1. Summary of measurements, showing the maximum stable current $I_{max}$(nA/cm) in a damaged chamber with Helium:Isobutane (80:20) gas before the additive, then with the additive shown, and then after the additive is removed. Cases that did not reach break down at the maximum attempted current are marked with a ">" sign. T is the training time to reach $I_{max}$. Some additives cure a damaged chamber, as indicated in the last column.

| Additive | (%) | Before | With Additive | | After | Cure? |
|---|---|---|---|---|---|---|
| | | $I_{max}$ | T(hr) | $I_{max}$ | $I_{max}$ | |
| Methylal | 4 | 0.3 | ~0 | >8 | | |
| | 2 | | ~0 | 3 | 0.4 | No |
| Propanol | 1.0 | 0.7 | ~0 | >12 | | |
| | 0.5 | | ~0 | >10 | | |
| | 0.25 | | ~0 | >13 | 0.2 | No |
| $H_2O$ | 0.35 | 0.4 | ~0 | >27 | | |
| | 0.18 | | ~0 | >9 | 0.5 | No |
| $O_2$ | 0.10 | 0.5 | 1.5 | >32 | >40 | Yes |
| | 0.05 | 0.4 | 2 | >29 | >16 | Yes |
| | 0.02 | 0.9 | 10 | >35 | >14 | Yes |
| $CO_2$ | 5 | 0.4 | 35 | >40 | >27 | Yes |
| $O_2 + H_2O$ (0.05%+0.35%) | | 0.4 | 40 | 10 | 3 | Partly |

A strange behavior already seen in Fig. 4 bottom is the rise and fall of the small pulse activity after the shutter is opened. This was seen for all gases with or without additives, for currents slightly below the break down threshold. Fig. 7 shows this effect for the Helium:Isobutane (80:20) gas.

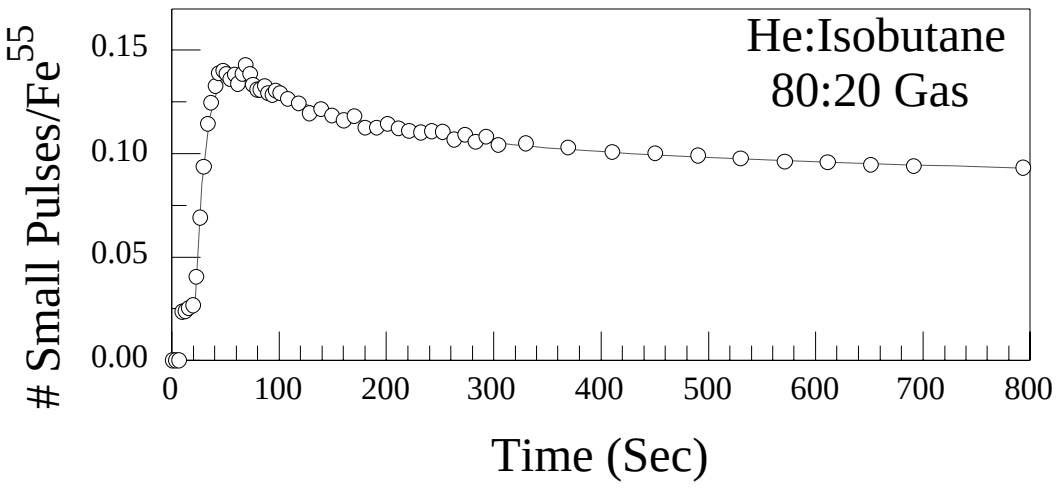

Figure 7. Small pulse activity at ionisation levels just below the break down threshold shows a rise and then a fall from a step increase in $Fe^{55}$ ionisation.

The increase in small pulses is readily explained by the Malter theory, in which an insulating deposit on the cathode collects the incoming positive ions and charges up to a point where the resulting high electric field draws off electrons from the insulator. As the charge and the electric field build up with time, the rate of emission of electrons also increases as observed, until Malter break down occurs or the charge build up is balanced by ohmic current flow through the insulator.

But the sudden reduction in small pulses after a build up period means that some other mechanism turns on which suppresses emission of electrons from the insulator. A likely candidate is heat. The positive ions gain energy when traversing the 19 KV/cm electric field at the cathode, and this energy is dissipated as heat in the collisions with the gas near the insulator or the insulating surface itself. As the insulator heats up, its effective resistance may fall. The charge build-up would then discharge more rapidly and the emission of electrons would decrease as observed.

## 4. Whiskers

After the test chamber was maximally damaged, the chamber was opened and the wires were viewed with a microscope. A few whiskers were found on both the anode and cathode wires. An example of a long whisker on a cathode wire is shown in Fig. 8. The whisker is white in color. Another cathode whisker (not shown) was of approximately equal size but was black in color.

The chamber was then operated with 0.05% oxygen and run until cured to the maximum source strength. Pictures were again taken of the same regions on the cathode wires. The white whisker was gone while the black one was still there as before.

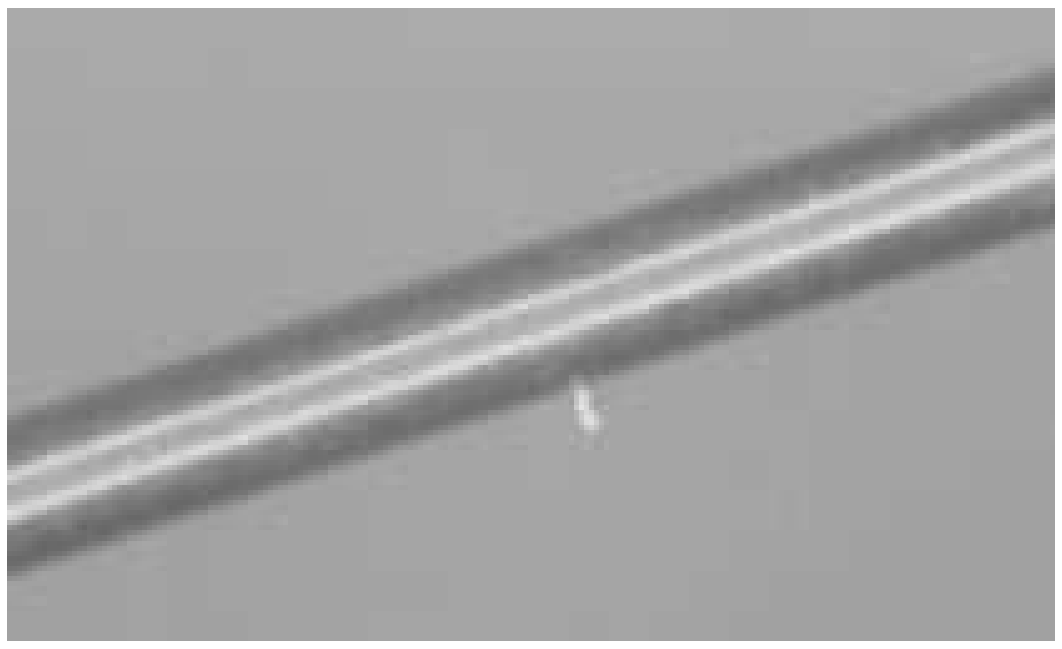

Figure 8. Picture of a whisker on a 120 um cathode wire after the chamber was purposely damaged by running at the maximum source strength and an elevated high voltage.

This suggests that the white whisker was a damaging element. It contributed to chamber break down and was removed by the oxygen cure. On the other hand, the black whisker did not cause break down at the highest ionisation level. However, this was a recent observation and there was no time to repeat it before the aging workshop. The white whisker could have fallen off the wire while transporting the chamber to the microscope room. This work will be repeated.

## 5. Conclusions

Damaged drift chambers that have excessive dark current or break down when running with Helium-Isobutane gas become immediately operable if water or alcohol is added to the gas. The best additives were found to be water (0.18-0.35%) or 2-Propanol (0.5-1%). Methylal (4%) is not as effective. When these additives are removed from the gas, the chamber returns to the damaged state.

Oxygen (0.02-0.05%) or carbon dioxide (5%) in the presence of high ionisation levels in the chamber allows a chamber to be slowly brought up to a high current level. These additives also cure a chamber from break down problems. When the additive is removed, the chamber can still operate at a high ionisation level (although it will start to age again if there is no additive present).

When operating a chamber just below the break down threshold, a new phenomenon is observed that decreases the number of small pulses and helps to prevent break down. It is postulated that this is due to ionic heating of the polymer deposits on the cathodes.

Whiskers are observed on the cathode wires of a damaged chamber.

## Acknowledgements

The author acknowledges the help from David Coupal during the initial planning of this project, from Al Odian for the many helpful discussions throughout the measurements, from Changguo Lu, and from Jim McDonald for constructing the test chamber and maintaining the apparatus.